# Magnetic and electrical anisotropy with correlation and orbital effects in dimerized honeycomb ruthenate Li$_2$RuO$_3$

Seokhwan Yun[1,2], Ki Hoon Lee[1,2], Se Young Park[1,2], Teck-Yee Tan[2], Junghwan Park[4], Soonmin Kang[1,2], D. I. Khomskii[5], Younjung Jo[3*] and Je-Geun Park[1,2*]

[1] *Department of Physics and Astronomy, Seoul National University, Seoul 08826, Korea*

[2] *Center for Correlated Electron System, Institute for Basic Science (IBS), Seoul 08826, Korea*

[3] *Department of Physics, Kyungpook National University, Daegu 41566, Korea*

[4] *Samsung SDI Co. Ltd., Yongin 17084, Korea*

[5] *II. Physikalisches Institut, Universität zu Köln, 50937 Köln, Germany*

* **Corresponding Author: jophy@knu.ac.kr & jgpark10@snu.ac.kr**

**Abstract**

Li$_2$RuO$_3$ undergoes a structural transition at a relatively high temperature of 550 K with a distinct dimerization of Ru-Ru bonds on the otherwise isotropic honeycomb lattice. It exhibits a unique herringbone dimerization pattern with an unusually large value of bond shrinkage, about ~ 0.5 Å. However, many questions still remain about its origin and its effect on the physical properties. In this work, using high-quality single crystals we investigated the anisotropy of resistivity ($\rho$) and magnetic susceptibility ($\chi$) to find a very clear anisotropy: $\rho_{c^*} > \rho_b > \rho_a$ and $\chi_b > \chi_a > \chi_{c^*}$. We also carried out density functional calculations for possible theoretical interpretations, and concluded that this anisotropic behavior is due to the correlation effects combined with the unique orbital structure and the dimerization of Ru 4$d$ bands.

## 1. Introduction

How a certain pattern of bonds forms for a given lattice underlies the fundamental physics and chemistry of the material concerned. The discovery by Kekulé of the resonating double-single carbon-carbon bonds for benzene is an outstanding case in point. A more modern example is the resonating valence bond state proposed by Anderson as a ground state for a triangular lattice with $S=1/2$ [1]. Therefore, it is a fundamental, important question to ask why a particular bond differs from others, in a seemingly equivalent environment. Another example is the work of Peierls, who discovered an instability in a one-dimensional lattice with one electron per ion, now known as the Peierls instability transition or dimerization [2]. As one moves to two-dimensionality or even 3D, the original argument of Peierls becomes perhaps less strict. However, when present with other factors such as orbital degrees of freedom, physics becomes more delicate and richer [3]. There are several examples in which the orbital degree of freedom triggers certain dimerization phenomena, a notable example among Ru oxides being $Tl_2Ru_2O_7$ [4].

$A_2MO_3$ (A=Li or Na, M=transition metal) is a promising candidate in the search for novel states originating from certain structural, electronic, and magnetic configurations on a honeycomb lattice. Many factors are expected to affect the competition among Kitaev physics, magnetism, and dimerization for the honeycomb lattice. On a microscopic level, they are governed by a number of transition metal $d$ electrons, the strength of spin-orbit coupling, the strength of correlation effects, the Hund's rule coupling, and the ionic radii of the A-site element [5,6]. As the relevant energy scales of these factors become more comparable to one another for $4d$ orbitals, the ground state of $4d$ transition metal oxides, in particular of ruthenates, becomes more sensitive to perturbation with the particular importance of the orbital degree of freedom [3,4,7]. One should also note that Ru has a moderate spin-orbit interaction of 75 meV as compared with its $5d$ counterparts.

Honeycomb lattice ruthenate $Li_2RuO_3$, with four $4d$ electrons in the $t_{2g}$ manifold, has attracted significant attention owing to the presence of an orbital-selective Peierls transition (OSPT) leading to the strongest tendency toward dimerization among the known $A_2MO_3$ systems [8,9,10]. It undergoes a strong dimerization transition below a quite high transition temperature (~540 K) as reported in [8]; it also displays a large direct overlap between $4d$ orbitals of the transition metal ions compared to other $Li_2MO_3$ structures. Most interestingly, the dimerized bonds between Ru atoms exhibit a herringbone pattern with a significant difference in length between the short and long bonds alternating along the bond direction, and at the same time it changes space group from C2/m to P2$_1$/m. It should be noted that this difference between the two short/long bonds is as large as ~ 0.5 Å, which is consistent with the strong direct bonding between two neighboring Ru atoms for the shorter bond. The short Ru-Ru bonds in $Li_2RuO_3$ are actually shorter than those in Ru metal. The local dimers of the short Ru-Ru bonds exhibit structural long-range order and form a valence bond solid (VBS) with a local spin singlet state [8,10]. In addition, at higher temperatures the dimers seem to survive locally [11], and the system exhibits a reduced local magnetic moment of $S = 1/2$, instead of the $S = 1$ [11] which would correspond to a typical $4d^4$ electron configuration at the octahedral site. Electronic structure calculations show that for each bond there is one pair of Ru $4d$ orbitals that overlap directly through $\sigma$-bonding. This bond is then responsible for the formation of the dimer with a singlet character [8]. We note that an OSPT-based dimer formation induced by orbital degeneracy was previously proposed to explain the above properties [9,11].

The directional orbital dependence is known to have a larger effect in the case of edge- or face-sharing octahedral geometries than in a corner-sharing geometry [3]. The valence electrons of Ru in the in-plane orbitals share the oxygen octahedral edges and form a strong $\sigma$-bond with bond strength comparable to the intra-atomic Hund's coupling energy. However, electrons in the orbitals that are orthogonal to the $Ru_2O_2$ dimer plane form weaker $\pi$- and $\delta$-bonds. The electronic structure from the $\pi$- and $\delta$-bonds would be significantly affected by the local dimerization. It is also expected that the contribution to the electrical and magnetic properties from the $\sigma$-bond participating in the dimer formation may be relatively small due to their strong binding energies. For instance, the electrons participating in the dimer bonds are located a few eV below the Fermi-level [10,11]. On the other hand, the electrons belonging to the weaker $\pi$- and $\delta$-bonds, which are located close to the Fermi-level, would more directly influence the physical properties of $Li_2RuO_3$. In particular, the bands occupied by these latter electrons are degenerate at the zone boundary owing to the lattice symmetry, the nonsymmorphic symmetry of $2_1$, which somehow went unnoticed in the previous calculations.

This degenerate band can be easily perturbed by the considerable spin-orbit coupling of Ru, forming a flat band; which itself is a very interesting observation of its own with potentially intriguing possibilities to explore.

It is worth noting that the electronic structure of $Li_2RuO_3$ is strongly modified near the Fermi-level by several factors: the orbital degree of freedom, the correlation effects, the spin-orbit coupling, and the strong dimerization. As the energy scales of each of them are more or less comparable to one another, it becomes increasingly more important to assess carefully what aspect of the physical properties of $Li_2RuO_3$ is driven by which one of them. To address this question, we investigated the electrical and magnetic anisotropies in details using high-quality single crystals of $Li_2RuO_3$. We also carried out density-functional band calculations to assist in interpreting the experimental results.

## 2. Experimental methods

$Li_2RuO_3$ single-crystals were synthesized by a self-flux growth method. The starting materials were $Li_2CO_3$ (99.995%, Alfa Aesar) and $RuO_2$ (99.95%, Alfa Aesar). Stoichiometric quantities of the materials plus an 8% excess of $Li_2CO_3$ were placed in an alumina crucible and heated sequentially at 600, 900, and 1000 °C for 24 h at each of the temperatures. The resulting powder was then pelletized and further subject to heat-treatment at 1100 °C for 48 h, which yielded shiny hexagonal crystals (a typical size of ~ 200 μm) as shown in the inset of Fig.1a. The quality and orientations of the obtained crystals were confirmed by X-ray diffraction (XRD) using a Rigaku XtalLab P200 (Mo target, averaged $K_\alpha$) and single crystal analysis using the WinPLOTER program [14]. Figure 1(c) shows the XRD patterns of the single crystal data along the c*-axis.

The resistivity was measured along the three principal crystallographic axes using a two-probe method due to the small sample size (Fig. 3a). The dimensions of the samples were as follows: for $\rho_a$ –[$w$(width)=125 μm, $l$(length)=151 μm, $h$(height)=47 μm], $\rho_b$ –[$w$=249 μm, $l$=200 μm, $h$=54 μm], $\rho_c$ –[$w$=220 μm, $l$=51 μm, $h$=221 μm]. The voltage applied between the two electrodes was kept below 0.2 V to avoid any possible charging effects that could arise from the high mobility of the Li$^+$ ions [15]. We used two different set-ups for our resistivity measurements to cover a wide temperature range: one with a cryostat covering the temperature range from 5 to 300 K and another one with a furnace covering the range from 300 to 650 K. We also measured the high-temperature resistivity along the b-axis with a 4-probe method to check the effect of contact resistance in the 2-probe data, thermal hysteresis between heating and cooling of sample, and the character of conductance in the high-temperature phase (Fig.3b). The dimensions of the sample are as follows: $w$=44.4 $\mu$m, $l$=200 $\mu$m, $h$=32.5 $\mu$m.

For the magnetic susceptibility measurements, we aligned approximately 250 pieces of the crystals with the total mass of ~ 1.091 mg along the $c$*-axis (perpendicular to the Ru honeycomb layer) using Kapton tape and stacked them in five layers (see the photo in the inset of Fig.4a). The susceptibility measurement was then performed from room temperature down to 2 K, in an applied magnetic field of 1 T parallel to the $c$*-axis and perpendicular to the $a$-axis, using a commercial magnetic property measurement system (MPMS3, Quantum Design). Then we calculated the ionic diamagnetic contribution of the susceptibility using the table in [16] and subtracted the calculated value from the measured susceptibility.

The magnetic anisotropy in the *ab*-plane was further measured with a torque magnetometer due to the small size of the single crystal. After checking the crystallinity and the crystallographic axes of the sample with single crystal XRD, we mounted it on the top of a piezo-resistive microcantilever and measured the magnetic torque along $\theta_{c*}$ with $\phi_{ba}$-rotation (Fig. 4c): $\theta_{c*}$ is the angle between the direction of the applied field and the $c$*-axis while $\phi_{ba}$ is the azimuthal angle in the *ab*-plane (see Fig.4c). All the measurements were performed using a commercial physical property measurement system (PPMS-9, Quantum Design) with a rotator.

We carried out the density functional theory(DFT) calculations using WIEN2k [17] with 12×6×12 k-points in the full Brillouin zone using the Tran-Blaha modified Becke Johnson (TB-mBJ) potential for exchange correlation [18,19]. TB-mBJ potential is known to give a better estimate of band gap than the standard functionals such as local density approximation (LDA) or generalized gradient approximation (GGA) functionals, without too much computational cost (for example, see [29]). The effect of spin-orbit coupling was included in the calculations (spin-orbit coupling strength of Ru is about 75 meV). The resistivity was then calculated with a BoltzWann module in Wannier90 to estimate the anisotropy of the resistivity [20,21]. BoltzWann uses a semi-classical Boltzmann transport equation to compute resistivity, which assumes a constant relaxation time approximation and a dispersion relation from the Wannierized tight-binding Hamiltonian. We then estimated the macroscopic magnetic susceptibility from the macroscopic susceptibility outputs provided by the NMR calculation module of WIEN2k [22,23]. The macroscopic magnetic susceptibility is calculated using the 2nd order perturbation theory on the DFT results by taking periodically modulated external magnetic fields with a long wavelength.

## 3. Results

Figure 1a shows the single crystal refinement result of the XRD data measured at room temperature. The number of peaks used for the refinement is approximately 1,700; the refined lattice parameters are a = 4.931 Å, b= 8.795 Å and c= 5.132 Å with a $\beta$ angle of 108.22 °. As reported in previous results, our result also shows a significant difference in the Ru-Ru bond lengths within the Ru honeycomb layer [8]. According to our analysis, the shortest Ru–Ru bond (red links in Fig.1b) length ($d_S$) is 2.571 Å, while the lengths of the other bonds (orange links in Fig.1b) are 3.048 Å ($d_I$) and 3.058 Å ($d_L$), where $(d_L - d_S)/d_S$ ~ 0.186: almost identical to the largest value reported previously [8].

Figure 2a shows the intensity of the (101) peak, one of the dimerization-related superlattice peaks, as a function of temperature up to 600 K. It clearly disappears above the transition temperature of approximately 550 K with a structural phase transition from P2$_1$/m to C2/m. Figure 2b shows the relative ratio of the two lattice parameters *b*/*a* as a function of temperature. We note that this ratio can be used as a quick quality check of the samples; it is sensitive to the disorder of dimers [8, 13, 24]. Table 1 displays the list of the b/a ratios reported in previous studies together with our own value: for our sample the ratio is found to be 1.784 at 300 K. Upon heating, it converges to $\sqrt{3}$ at 600 K with a uniform Ru-Ru bond length [8] and the hexagons forming the Ru honeycomb layers become almost regular at higher temperatures (inset in Fig.2b).

To investigate the effects of both correlation and spin-orbit coupling on the dimerized state, we examined the anisotropy of physical properties using single crystal samples: both the correlation and the spin-orbit coupling are expected to induce nontrivial anisotropy in physical properties. The resistivity curves in Fig.3c show the phase transition around 550 K with just such clear anisotropic behavior. The resistivity along the $c^*$-axis is the largest over the whole temperature range probably because Li layers separate the honeycomb layers along this direction. Of particular interest is an in-plane anisotropy: the $b$-axis resistivity is larger than the $a$-axis resistivity. This in-plane anisotropy implies that an inter-dimer electronic hopping along the $b$-axis ($d_I$ and $d_L /d_S$ in Fig.1b) is smaller than that along the a-axis ($d_L$ and $d_S$ in Fig.1b). This resistivity ratio $\rho_b/\rho_a$ is found to be about 2 above the phase transition, but it increases with decreasing temperature and reaches around 10 at 5 K. This large in-plane anisotropy indicates directly the strong directional anisotropy of the hopping integrals in the dimerized phase, which is most likely due to the orbital degree of freedom of the Ru 4$d$ bands. We also measured $\rho_b(T)$ with a 4-probe method to confirm the negligible effect of contact resistance; the temperature dependence and the value of the resistivity are almost the same as those of the 2-probe result (Fig.3d). These data exhibit a clear thermal hysteresis and insulating behavior in the high-temperature phase.

We also show an Arrhenius plot of the resistivity data in the inset of Fig.3c. For all three crystallographic directions, the curves are well fitted with a straight line with an energy gap of approximately 0.15 eV in the high-temperature phase. However, they do not follow the activation formula in the low-temperature state with the herringbone pattern of dimerized bonds. To check whether this low-temperature behavior can be explained by topological insulating behavior as suggested for SmB$_6$ [25], we have used VASP2trace to examine the band topology and found that all the bands are topologically trivial in Li$_2$RuO$_3$ [26]. Thus, we believe that this low-temperature flattening behavior is more likely to be extrinsic, probably due to Li defects.

To examine this anisotropic behavior further, we carried out magnetic susceptibility measurements. The susceptibility curves shown in Fig.4a are almost temperature-independent below the transition temperature due to the singlet formation of the 4$d$-electrons in the Ru dimers [8]. The typical up-turn behavior at low temperatures, most probably originating from paramagnetic impurities, is observed independently of direction. But what is most remarkable is that the low-temperature susceptibility has a large van Vleck paramagnetic contribution of $4\sim6 \times 10^{-4}$ emu/mol, which is much larger than the ionic diamagnetic contribution of $-5.6\times 10^{-5}$ emu/mol [16]. We also note that the $c^*$-axis susceptibility is smaller than that along the in-plane direction. This anisotropy in the susceptibility is consistent with the reported data in Refs. [8, 13, 27], but not that of Ref. [24]. Note that the ratio of those two values ($=\chi_\perp/\chi_\parallel$) is 0.70 in our data, similar to that of [27]; in contrast, $\chi_\perp$ was reported to be larger than $\chi_\parallel$ in [24].

As shown in Fig.4b and 4c, the magnetic easy axis can be uniquely identified from a complete angular dependence of the torque measurements, $\tau(\theta_{c*})$. The amplitude of the $\sin2\theta_{c*}$-dependence is proportional to the principal components of magnetic anisotropy $\alpha_{ij}$. Figure 4b shows that $\tau(\theta_{c*})$ at different azimuthal angles $\phi_{ba}$, and $\tau(\theta_{c*})$ can be fitted with a sinusoidal function with a period of $\pi$. From these torque measurement data, we find that the susceptibility in any in-plane direction is larger than that along the out-of-plane direction, confirming our susceptibility measurement results shown in Fig. 4a. Figure 4c shows the fitted amplitude of the data with a $b$-to-$a$ rotation. The difference between $\chi_{ab}$ and $\chi_{c*}$ depends on the azimuthal angle; it is the largest along the $b$-axis and smallest along the $a$-axis. The ratio defined as $\alpha_{bc*}/\alpha_{ac*} = (\chi_b - \chi_{c*})/(\chi_a - \chi_{c*})$ is found to be approximately 3.5 at 10 K.

## 4. Discussion and analysis

To understand the anisotropic behavior observed in both resistivity and susceptibility, we carried out DFT calculations with the effect of spin-orbit coupling included (see Fig.5a). An important point worth noting here is that when we used the standard potential (GGA or LDA) for our DFT calculation, we could not open a band gap. Only when we used the TB-mBJ potential did we succeed in inducing an indirect band gap of 170 meV: whose value is more or less consistent with the experimental values. This dependence of the band gap on the potentials used for the DFT calculations implicitly implies that the Coulomb $U$ plays an important role in realizing the insulating phase, which is embedded in the TB-mBJ potential.

With the band structures producing the correct value of the band gap, we then calculated both the resistivity and susceptibility results using the modules embedded in the WIEN2k code. According to our calculations done without spin-orbit coupling, we obtained the following values: $\chi_a$ = 1.83, $\chi_b$ = 2.34, and $\chi_c$ = 1.21, all in units of $10^{-4}$ emu/mol [22,23]. Remarkably, these calculations not only give the correct anisotropy but also give values of the same order of magnitude as experimentally found. And we also calculated the resistivity based on the same band structure using a semi-classical Boltzmann approach within Wannier90 [20,21]. The calculation was intended to capture the thermally excited charge carrier contribution under the constant relaxation time approximation. In this resistivity calculation, we succeeded in getting the correct anisotropy of out-of-plane and in-plane resistivity as one can see in Fig.5c. Interestingly, one can change this anisotropy in the resistivity by shifting the Fermi level by 0.1 eV with more electron doping. But we note that our calculations failed to produce the correct in-plane anisotropy of the resistivity. Experimentally, we found that the a-axis resistivity is smaller than the b-axis while our calculations suggest a b-axis resistivity smaller than a-axis. The origin of this discrepancy in the resistivity anisotropy is unclear at the moment. We guess that it may be due to the considerable anisotropic band renormalization due to the correlation effects, which goes beyond the scope of our attempted DFT calculations and may be a subject of further theoretical studies.

Another point worth noting is the band degeneracy along some specific momentum directions: the Z-D and E-Z-C2-Y2 directions, as shown in Fig.5a. Without the spin-orbit coupling, it is perfectly degenerate and becomes slightly split with a spin-orbit coupling of 75 meV. This degeneracy is protected by the nonsymmorphic symmetry of the low-temperature phase of $P2_1/m$. This degenerate and almost flat band gives rise to a large density of states just below the Fermi level: our Hall experiment shows that $Li_2RuO_3$ is intrinsically *n*-type. Thus, with some control of the Fermi level, such as gating experiments, one may be able to control the ground state - an interesting direction for future research.

Of further note, the metallic solution (e.g. in LDA or GGA methods) does not correctly reproduce the experimentally measured anisotropy in the physical properties. On the other hand, as we discussed above, our calculations with the Hubbard's U give the reasonable description of the observed anisotropy. This is the argument in favor of the Coulomb correlations (U) playing important role in the behavior of $Li_2RuO_3$.

To summarize, we found the clear experimental evidence of a strong anisotropy in both resistivity and susceptibility data for single crystals of $Li_2RuO_3$. Using theoretical studies, we verified that the anisotropy in the susceptibility is reproducible with the DFT calculations with the TB-mBJ potential, indicating the importance of correlation effects.

**Acknowledgements:** We would like to thank Kisoo Park, Yukio Noda, Igor Mazin, Giniyat Khaliullin, and George Jackelli for useful discussion. Special thanks should be given to Matt Coak for his careful reading of the manuscript and comments. Y. J. Jo was supported by National Research Foundation, Korea (NRF-2016R1A2B4016656, NRF-2018K2A9A1A06069211) and the work of D. Khomskii was funded by the Deutsche Forschungsgemeinschaft (DFG) - Project number 277146847 - CRC 1238. The work at the IBS CCES and SNU was supported by the Institute of Basic Science (IBS) in Korea (Grants No. IBS-R009-G1).

| | References | | b/a | $(b/a-\sqrt{3})/\sqrt{3}$ (%) |
|---|---|---|---|---|
| Calculation | [11] | VASP | 1.789 | 3.260 |
| Powder | [8] | $Li_2RuO_3$ | 1.785 | 3.044 |
| | [12] | LRO1 | 1.780 | 2.779 |
| | | LRO2 | 1.784 | 2.982 |
| | | LRO3 | 1.780 | 2.740 |
| | | LRO4 | 1.777 | 2.581 |
| | | LRO5 | 1.776 | 2.561 |
| | | LRO6 | 1.767 | 2.039 |
| | [13] | A | 1.774 | 2.438 |
| | | B | 1.771 | 2.261 |
| | | C | 1.781 | 2.836 |
| | | D | 1.782 | 2.855 |
| | | E | 1.782 | 2.893 |
| | | F | 1.784 | 2.975 |
| | | G | 1.785 | 3.042 |
| | | H | 1.785 | 3.061 |
| | [27] | $Li_2RuO_3$ | 1.781 | 2.799 |
| | | $(Li_{0.95}Na_{0.5})_2RuO_3$ | 1.778 | 2.669 |
| | [24] | $Li_2RuO_3$ | 1.785 | 3.044 |
| Single Crystal | [24] | $Li_2RuO_3$(P) | 1.766 | 1.976 |
| | | $Li_2RuO_3$(C) | 1.744 | 0.671 |
| | This work | $Li_2RuO_3$ | 1.784 | 2.977 |

Table 1. Summary of b/a parameter taken after several references.

**Figure captions**

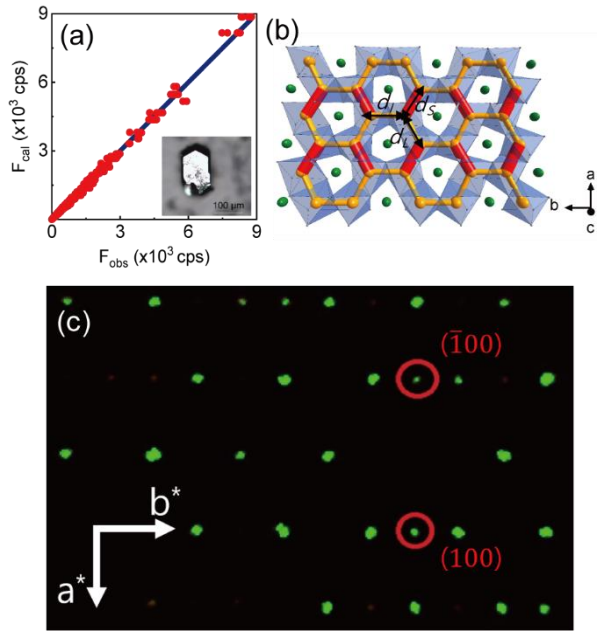

Fig. 1. (a) Single crystal refinement results of the $Li_2RuO_3$ single-crystal. The inset shows a typical hexagonal single-crystal. (b) $Li_2RuO_3$ at room temperature, viewed along the perpendicular direction to the Ru honeycomb layer in the *ab*-plane. The yellow and green spheres represent the Ru and Li ions, respectively. The blue polygons represent the oxygen octahedrons. There are two unequal Ru–Ru bonds, i.e. dimerized bonds (red) and two other bonds (yellow) with similar lengths. (c) X-ray diffraction image in the (hk0) plane of single-crystal $Li_2RuO_3$ with no sign of twinning.

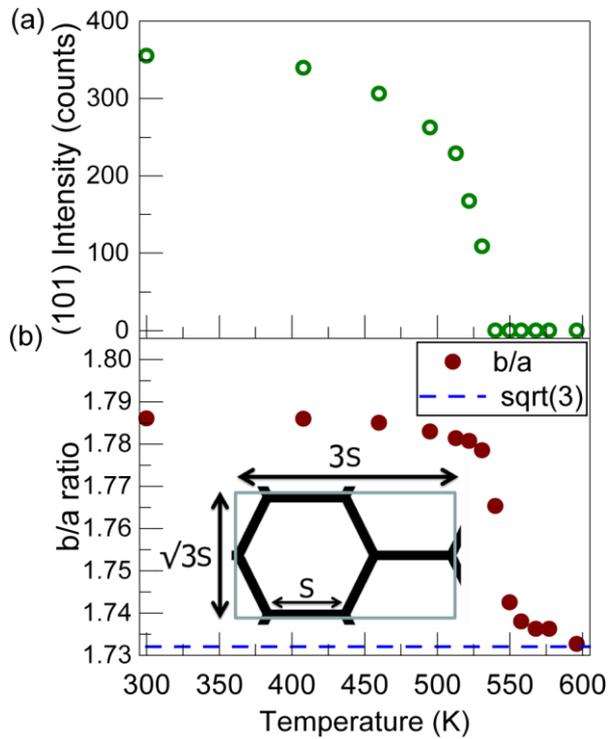

Fig. 2. Temperature dependence of (a) the intensity of the (101) peak and (b) the $b/a$ ratio of the lattice parameters. The blue dashed line represents the value of $b/a \sim \sqrt{3}$, a value found for the honeycomb structure with an almost ideal honeycomb lattice. The inset shows an illustration of the ideal hexagonal honeycomb structure.

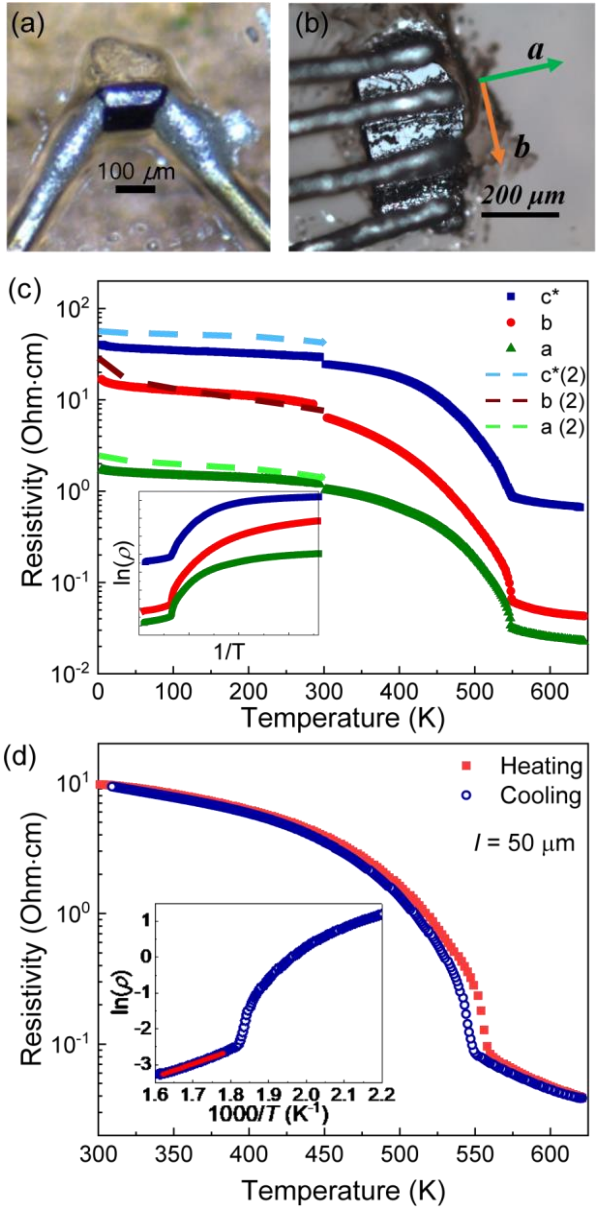

Fig. 3. The photographs of (a) 2-probe contact and (b) 4-probe contact on the single-crystal $Li_2RuO_3$ (c) Resistivity of the $Li_2RuO_3$ single-crystal as a function of temperature in the range of 5–650 K, along the *a*-(green), *b*-(red), and *c**-(blue) crystal axes. The dashed lines in the range of 5–300 K are for the data taken on another samples to check the reproducibility of the results. The inset shows the Arrhenius plots of the resistivity curves from 300 to 650 K. (d) 4-probe resistivity measurement along the b-axis in the range of 300-630 K, with heating (red) and cooling (blue). The inset shows the Arrhenius plot of the cooling curves.

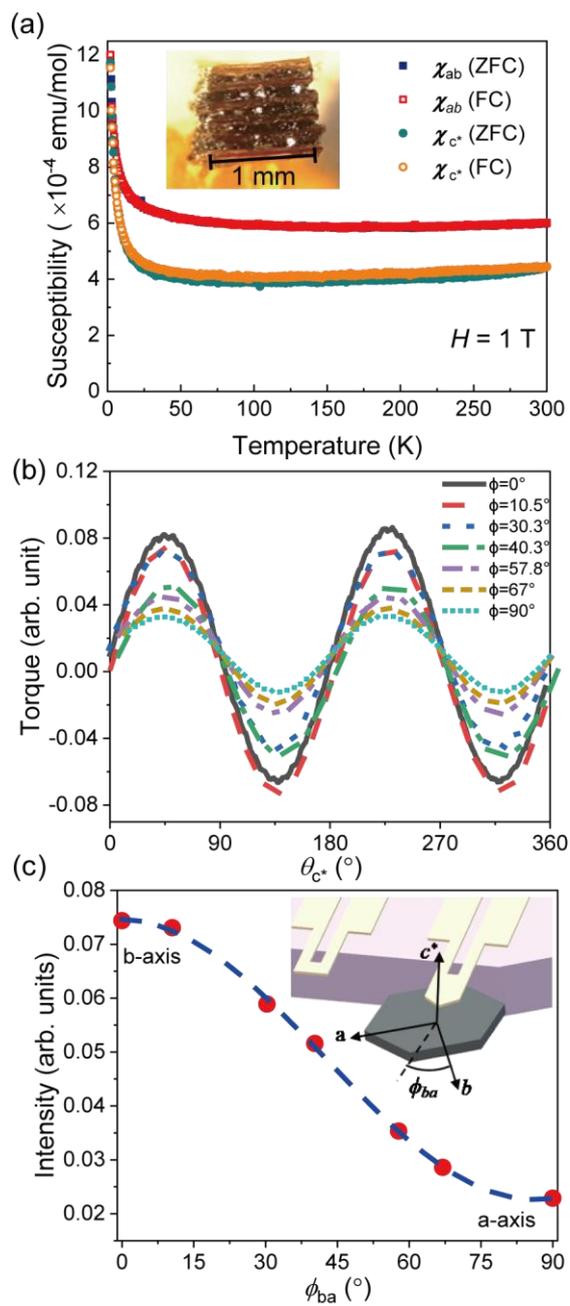

Fig. 4. (a) Susceptibilities of the *c**-axis-aligned single-crystals as a function of temperature in the range of 2–300 K, along the out-of-plane ($\chi_{c^*}$, circle marks) and in-plane ($\chi_{ab}$, square marks) directions; the inset shows the sample used for the measurement. (b) Angular-dependent torque measurement at fixed $\phi$ angles from $\phi_{ba} = 0°$ (*b*-axis) to $\phi_{ba} = 90°$ (*a*-axis). (c) Fitted amplitudes from the data with *ab*-rotation. The inset illustrates the crystal axes and rotating angles of the sample.

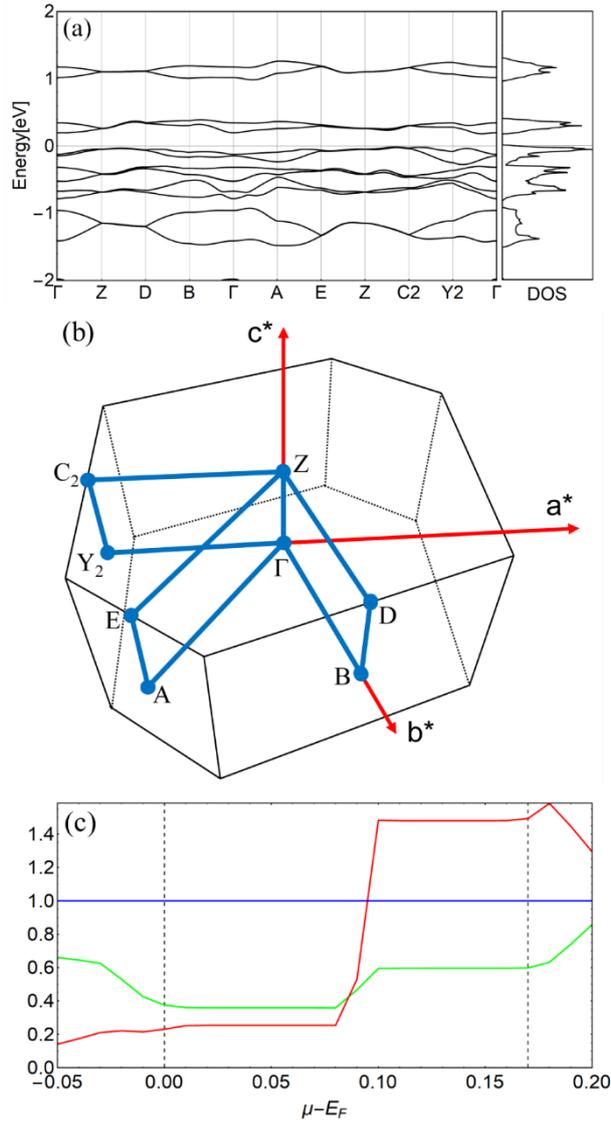

Fig. 5. (a) DFT band structure: upper six bands are from anti-bonding states and lower six bands are from bonding states. (b) Various points in the first Brillouin zone of the $Li_2RuO_3$ (c) Resistivity divided by $\rho_{c^*}$ at 100 K with varying chemical potential. The green line represents $\rho_a/\rho_{c^*}$ while the red line $\rho_b/\rho_{c^*}$.